\documentclass[aps, prd, floatfix, twocolumn, superscriptaddress, nofootinbib]{revtex4-1}

\usepackage{latexsym}
\usepackage{amsmath}
\usepackage{amssymb}
\usepackage{amsfonts}
\usepackage{mathrsfs} 

\usepackage{dsfont}
\usepackage{slashed}
\usepackage{bm}
\usepackage{urwchancal}

\usepackage[vcentermath]{youngtab}

\usepackage{color}
\definecolor{purple}{rgb}{0.5,0,0.5}
\definecolor{blue}{rgb}{0.0,0,0.9}
\definecolor{prdblue}{rgb}{0.133,0.118,0.498}
\usepackage{xcolor}

\usepackage{enumitem}
\usepackage{multirow}
\usepackage{subfigure}
\usepackage{textcomp}
\usepackage{dcolumn}
\usepackage{longtable}

\usepackage{supertabular} 
\usepackage{placeins}
\usepackage{epsfig}
\usepackage{graphicx}

\usepackage{natbib}

\begin{document}


\title{The bosonic algebraic approach applied to the $[QQ][\bar{Q}\bar{Q}]$ tetraquarks}

\author{A. J. Majarshin}
\email[]{jalili@nankai.edu.cn}
\affiliation{School of Physics, Nankai University, Tianjin 300071, P.R. China}

\author{Yan-An Luo}
\email[]{luoya@nankai.edu.cn}
\affiliation{School of Physics, Nankai University, Tianjin 300071, P.R. China}

\author{Feng Pan}
\email[]{daipan@dlut.edu.cn}
\affiliation{Department of Physics, Liaoning Normal University, Dalian 116029, P.R. China}
\affiliation{Department of Physics and Astronomy, Louisiana State University, Baton Rouge, LA 70803-4001, USA}

\author{J. Segovia}
\email[]{jsegovia@upo.es}
\affiliation{Dpto. Sistemas F\'isicos, Qu\'imicos y Naturales, Univ. Pablo de Olavide, 41013 Sevilla, Spain}

\date{\today}

\begin{abstract}
The exact eigenenergies of the $T_{4c}=[cc][\bar{c}\bar{c}]$, $T_{4b}=[bb][\bar{b}\bar{b}]$, and $T_{2[bc]}=[bc][\bar{b}\bar{c}]$ tetraquarks are calculated within the extended transitional Hamiltonian approach, in which the so-called Bethe \emph{ansatz} within an infinite-dimensional Lie algebra is used. 
We fit the parameters appearing in the transitional region from phenomenology associated with potential candidates of tetraquarks. The rotation and vibration transitional theory seems to provide a better description of heavy tetraquarks than other attempts within the same formalism.
Our results indicate that the pairing strengths are large enough to provide binding; an extended comparison with the current literature is also performed.
\end{abstract}


\maketitle


\section{Introduction}
\label{sec:introduction}

A system of interacting bosons is a well-studied problem. Having its roots in the Bose-Einstein condensates~\cite{leg, pet, ber}, the framework has been applied to studies of nuclear and molecular structure~\cite{arim, oss91, oss93}, and examples of algebraic methods applied to hadron physics can be found in Refs.~\cite{f1, f2, f3, b1, b2}. We have recently applied the interacting boson approximation proposed by Arima and Iachello~\cite{arima75}, which includes two types of bosons (\emph{s} and \emph{d}-bosons), to the computation of wave functions in an interacting $sl$ many-body boson system~\cite{pan2002}.

Quarks can combine to form hadrons such as mesons (quark-antiquark pair) and baryon (three-quarks). Within an algebraic framework, the spectrum of hadrons began to be studied with the seminal work of Iachello in 1989~\cite{f1}. He was also able to elucidate some features about the structure of mesons and baryons, and the emergence of general patterns. An extension of the interacting boson approximation for studying eigenenergies of mesons in the $U(4)$ model was proposed by \emph{et al.} in 2006~\cite{pan2006}. The mass spectra of $Q\bar{Q}$ mesons, with $Q$ either $c$ or $b$ quark, has been recently discussed by the $U(3)\rightarrow O(4)$ transitional theory~\cite{f1, f2, pan2006}. Herein, we want to extend this model to multi-quark states~\cite{jaf}, in particular, tetraquarks with only heavy-quark content. In an extension of the $sl$ boson system, the largest dynamical symmetry group is generated by $s$ and $l$ $(l=Q,\,\bar{Q},\ldots)$ boson operators. We examine a similar Hamiltonian, based on $SU(1,1)$ algebraic technique~\cite{pan2002, aj17, aj181, aj182} and in a $sl$ boson system to describe the masses of the $[QQ][\bar{Q}\bar{Q}]$ tetraquarks. Our predictions will provide a new solvable model in hadron physics. We shall show that the masses of the $[QQ][\bar{Q}\bar{Q}]$ tetraquarks are sensible to the vector quark pairing strengths.

Fully-heavy tetraquarks have recently received considerable attention, both experimentally and theoretically. On the experimental side, it is thought that all-heavy tetraquark states will be very easy to spot because their masses should be far away from the typical mass regions populated by both conventional heavy mesons and the XYZ states discovered until now~\cite{Tanabashi:2018oca}. A search for deeply bound $bb\bar b \bar b$ tetraquark states at the LHC was motivated by Eichten \emph{et al.} in Ref.~\cite{Eichten:2017ual}, and it was carried out by the LHCb collaboration~\cite{Aaij:2018zrb} determining that no significant excess is found in the $\mu^+\mu^-\Upsilon(1S)$ invariant-mass distribution. Note, however, that a search for exotic mesons at the CMS experiment reported a potential candidate of a fully bottom tetraquark $T_{4b}=[bb][\bar{b}\bar{b}]$ around 18-19 GeV~\cite{dur}. On the other hand, the LHCb collaboration has recently released in Ref.~\cite{1804391} a study of the $J/\psi$-pair invariant mass spectrum finding a narrow peak and a broad structure which could originate from hadron states consisting of four charm quarks. 

From the theoretical side, we find fully-heavy tetraquark computations based on phenomenological mass formulae~\cite{Karliner:2016zzc, Berezhnoy:2011xn, Wu:2016vtq}, QCD sum rules~\cite{Chen:2016jxd, Wang:2017jtz, Wang:2018poa, Reinders:1984sr}, QCD motivated bag models~\cite{Heller:1985cb}, NR effective field theories~\cite{Anwar:2017toa, Esposito:2018cwh}, potential models~\cite{Ader:1981db, Zouzou:1986qh, Lloyd:2003yc, Barnea:2006sd, Richard:2018yrm, Richard:2017vry, Vijande:2009kj, Debastiani:2017msn, Liu:2019zuc, Chen:2019dvd, Chen:2019vrj, Chen:2020lgj, Wang:2019rdo, Yang:2020rih}, non-perturbative functional methods~\cite{Bedolla:2019zwg}, and even some exploratory lattice-QCD calculations~\cite{Hughes:2017xie}. Some works predict the existence of stable $QQ\bar Q\bar Q$ ($Q=c$ or $b$) bound states with masses slightly lower than the respective thresholds of quarkonium pairs (see, for instance, Refs.~\cite{Chen:2016jxd, Anwar:2017toa, Karliner:2016zzc, Berezhnoy:2011xn, Wang:2017jtz, Wang:2018poa, Debastiani:2017msn, Esposito:2018cwh}. In contrast, there are other studies that predict no stable $cc\bar c\bar c$ and $bb\bar b\bar b$ tetraquark bound states because their masses are larger than two-quarkonium thresholds (see, \emph{e.g.}, Refs.~\cite{Ader:1981db, Lloyd:2003yc, Richard:2018yrm, Wu:2016vtq, Hughes:2017xie}). To some extent, a better understanding of the  mass locations of fully-heavy tetraquark states would be desirable, if not crucial, for our comprehension of their underlying dynamics and their experimental hunting.


\section{Theoretical method}
\label{sec:theory}

Within this framework, diquark clusters must be assumed in order to describe a tetraquark system. According to this, a tetraquark
\begin{equation}
T = Q_1 Q_2 \bar{Q}_3 \bar{Q}_4 \,,
\end{equation}
contains two point-like diquarks and we extend the interacting boson model to multi-level pairing considering algebraic solutions of an $sl$-boson system~\cite{pan2002}. Note that the dynamical symmetry group is generated by $s$ and $l$ operators, where $l$ can be the configuration of the multiquark states. In the Vibron Model, elementary spatial excitations are scalar $s$-bosons with spin and parity $l^\pi=0^+$ and vector $l$-bosons with spin and parity $l^\pi=1^-$. In the finite-dimensional $SU(1,1)$ algebra, we have the generators, which satisfies the following commutation relations
\begin{subequations}
\begin{align}
\label{2a}
[Q^{0}(l),Q^{\pm}(l)] &= \pm Q^{\pm}(l) \,, \\
\label{2b}
[Q^{+}(l),Q^{-}(l)] &= -2Q^{0}(l).
\end{align}
\end{subequations}
Now, we apply the affine $\widehat{SU(1,1)}$ algebra for $U_{s}(1) \otimes U_{Q_1Q_2}(3) \otimes U_{\bar{Q}_3\bar{Q}_4}(3) \otimes U_{J}(3) - SO(10)$ transitional Hamiltonian. It is important to note that $Q_1Q_2\bar Q_3\bar Q_4$ responds, respectively, to $T_{4c}=[cc][\bar{c}\bar{c}]$, $T_{4b}=[bb][\bar{b}\bar{b}]$ and $T_{2bc}=[bc][\bar{b}\bar{c}]$ systems; and also that the quasi-spin algebras have been explained in detail in Refs.~\cite{pan2002, aj17, aj181, aj182}.

By using Eqs.~\eqref{2a} and~\eqref{2b} as generators of the $SU^{l}(1,1)$-algebra for tetraquarks, we have
\begin{subequations}
\begin{align}
Q^{+}({l}) &= \frac{1}{2} \, l^{\dag} \cdot l^{\dag} \,, \\ 
Q^{-}({l}) &= \frac{1}{2} \, {\tilde{l}} \cdot {\tilde{l}} \,, \\ 
Q^{0}({l}) &= \frac{1}{2} \, \left( l^{\dag} \cdot {\tilde{l}} + \frac{2l+1}{2} \right) \,,
\end{align}
\end{subequations}
where $l^{\dag}$ is the creation operator of an \emph{l}-boson constituying the tetraquark, and $\tilde{l}_\nu=(-1)^{\nu}l_{-\nu}$.

It is accepted that the basis vectors of $U(2l+1)\supset SO(2l+1)$ and $O(2l+2)\supset O(2l+1)$ are simultaneously the basis vectors of $SU(1,1)^l \supset U(1)_s ^l$ and $SU(1,1)^{sl} \supset U(1)_s ^{sl}$, respectively. Their complementary relation for tetraquark states can be expressed as
\begin{align}
|Q_1,Q_2,\bar{Q}_3,&\bar{Q}_4,N; n_l\, \nu_l\,, n_\Delta JM\rangle = \nonumber \\
&
=|Q_1,Q_2,\bar{Q}_3,\bar{Q}_4,N; \kappa_l\, \mu_l\,, n_\Delta JM\rangle \,,
\end{align}
with $\kappa_l=\frac{1}{2}\nu_l+\frac{1}{4}(2l+1)$ and $\mu_l=\frac{1}{2}n_l+\frac{1}{4}(2l+1)$; and where $N$, $n_l$, $\nu_l$, $J$ and $M$ are quantum numbers of $U(N)$, $U(2l+1)$, $SO(2l+1)$, $SO(3)$ and $SO(2)$, respectively. The quantum number $n_\Delta$ is an additional one needed to distinguish different states with the same $J$. However, the pairing models of multi-level are also characterized by an overlaid $U(n_{1}+n_{2}+\ldots)$ algebraic structure which has been described in detail in, for istance, Refs.~\cite{pan2002, pan2006, aj181, aj182}. This is to say either
\begin{align}
\mathop{U(10)}\limits_N \supset \mathop{U(9)}\limits_{n_l}
\end{align}
or
\begin{align}
\mathop{U(10)}\limits_N &\supset \mathop{SO(10)}\limits_{\nu_l} \supset \mathop{SO(9)}\limits_{\nu_l} \nonumber \\ 
&
\supset \mathop{SO(3)}\limits_s  \otimes \mathop{SO(3)}\limits_{Q_1Q_2}  \otimes \mathop{SO(3)}\limits_{\bar{Q}_3 \bar{Q}_4}  \otimes \mathop {SO(3)}\limits_J \,.
\end{align}

Affine Lie algebras are famous among the infinite-dimensional Lie algebras and have widespread applications because of their representation theory. We know that affine Lie algebra is far richer than that of finite-dimensional simple Lie algebras. Hence, in contrast to Eqs.~\eqref{2a} and~\eqref{2b}, the operators under the corresponding $SU(1,1)$ irreducible representations  satisfy the following commutation relations:
\begin{subequations}
\begin{align}
\label{3a}
[Q^{0}_{m}(l),Q^{\pm}_{n}(l)] &= \pm Q^{\pm}_{m+n}(l) \,, \\
\label{3b}
[Q^{+}_{m}(l),Q^{-}_{n}(l)] &= -2Q^{0}_{m+n+1}(l).
\end{align}
\end{subequations}
According to the definitions, $Q^{\Omega}_{m}$, with $\Omega=0,\pm$ and $m=0,\,\pm1,\,\pm2,\ldots$ generate the affine Lie algebra without central extension. The infinite dimensional $\widehat{SU(1,1)}$ Lie algebra defined by
\begin{subequations}
\begin{align}
\label{4a}
Q_{n}^{\pm} &= c_{Q_1}^{2n+1} Q^{\pm}(Q_1) + c_{Q_2}^{2n+1} Q^{\pm}(Q_2) + c_{\bar{Q}_3}^{2n+1} Q^{\pm}(\bar{Q}_3) \nonumber \\
&
+ c_{\bar{Q}_4}^{2n+1} Q^{\pm}(\bar{Q}_4), \\
\label{4b}
Q_{n}^{0} &= c_{Q_1}^{2n} Q^{0}(Q_1) + c_{Q_2}^{2n} Q^{0}(Q_2) + c_{\bar{Q}_3}^{2n} Q^{0}(\bar{Q}_3) \nonumber \\
&
+ c_{\bar{Q}_4}^{2n} Q^{0}(\bar{Q}_4) \,,
\end{align}
\end{subequations}
where  $c_{Q}$'s are real-valued control parameters for tetraquarks, and $n$ can be taken to be $1,\,2,\,3,\,\ldots$

The lowest weight state of fully-heavy tetraquarks should satisfy $Q^{-}(l)|lw\rangle=0$. Then, we define $|lw\rangle$ by the following expression:
\begin{align}
|lw\rangle=|Q_1,Q_2,\bar{Q}_3,\bar{Q}_4,{N};\kappa_{l}\,\mu_{l}, n_\Delta JM \rangle,
\end{align}
where $ N=2k+ \nu_{Q_1}+ \nu_{Q_2}+ \nu_{\bar{Q}_3}+ \nu_{\bar{Q}_4}$.
Hence, we have
\begin{equation}
Q_{n}^{0} |lw\rangle =\Lambda_{n}^{l} |lw\rangle , \,\,\, \Lambda_{n}^{l}=\sum_l c_{l}^{2n}\frac{1}{2} \left(n_{l}+\frac{2l+1}{2} \right).
\end{equation}

It is apparent that the system is in the vibrational $U(9)$ and rotational $SO(10)$ transition region as the pairing strengths, $c_l$, vary continuously within the closed interval $[0,c_l]$. The quantum phase transition occurs in the all-heavy tetraquark pairing model. The $U(9)$ limit is fulfilled by ${c_{Q_1} = c_{Q_2} = c_{\bar{Q}_3} = c_{\bar{Q}_4} = 0}$ where as the $SO(10)$ limit occurs when ${c_{Q_1} = c_{Q_2} = c_{\bar{Q}_3} = c_{\bar{Q}_4} = 1}$. In our calculation, we have extracted different values for the control parameters between $U(9)$ and $SO(10)$ limits, \emph{viz.} $c_{Q_i} \in [0,1]$ with $i=1,\ldots,4$.

The total Hamiltonian is represented in terms of the Casimir operators  ${\hat C_2}$ by branching chains. The two first terms of the Hamiltonian, $Q_{0}^{+} Q_{0}^{-}$ and $Q_{1}^{0}$, are related to $SU(1,1)$ algebra and the remaining ones are constant according to Casimirs. In duality relation for tetraquarks, the irreducible representations reduce the quasi-spin algebra chains~\eqref{4a} and~\eqref{4b} as well, and the labels for the chains are connected through the duality relations. By employing the generators of algebra $SU(1,1)$, the proposed Hamiltonian for heavy tetraquark pairing model is
\begin{align}
\label{a01}
\hat H &= g \, Q_0^+ Q_0^- + \alpha \, Q_1^0 + \beta \, \hat{C}_2(SO(9)) \nonumber \\
&
+ \gamma_1 \, \hat{C}_2(SO(3)_s) + \gamma_2 \, \hat{C}_2(SO(3)_{Q_1Q_2}) \nonumber \\
&
+ \gamma_3\, \hat{C}_2(SO{(3)_{\bar Q_3\,\bar Q_4}}) + \gamma \, \hat{C}_2(SO{(3)_J}),
\end{align}
where $g$, $\alpha$, $\beta$, $\gamma _1$, $\gamma _2$, $\gamma _3$, and $\gamma$ are real-valued parameters.

To find the non-zero energy eigenstates with $k$-pairs, we exploit a Fourier Laurent expansion of the eigenstates of Hamiltonians which contain dependences on several quantities in terms of unknown $c$-number parameters $x_i$, and thus eigenvectors of the Hamiltonian for excitations can be written as
\begin{align}
&
|k;\nu_{Q_1}\nu_{Q_2}\nu_{\bar{Q}_3}\nu_{\bar{Q}_4} n_\Delta JM \rangle = \sum_{n_{i}\in Z} a_{n_{1}n_{2} \ldots n_{k}} \nonumber \\
&
=  x_{1}^{n_{1}} x_{2}^{n_{2}} x_{3}^{n_{3}} \ldots x_{k}^{n_{k}}Q_{n_{1}}^{+} Q_{n_{2}}^{+} Q_{n_{3}}^{+} \ldots Q_{n_{k}}^{+}|lw\rangle \,,
\label{wave}
\end{align}
and
\begin{align}
\label{quasi}
Q _{n_{i}}^+ &= \frac {c_{Q_1}}{1-c_{Q_1}^{2} x_{i} } Q^{+} (Q_1)+\frac {c_{Q_2}}{1-c_{Q_2}^{2} x_{i} } Q^{+} (Q_2) \nonumber \\
&
+ \frac {c_{\bar{Q}_3}}{1-c_{\bar{Q}_3}^{2} x_{i} } Q^{+} (\bar{Q}_3)+\frac {c_{\bar{Q}_4}}{1-c_{\bar{Q}_4}^{2} x_{i} } Q^{+}(\bar{Q}_4) \,.
\end{align}
The $c$-numbers, $x_i$, are determined through the following set of equations
\begin{equation}
\frac{\alpha}{x_{i}}=\sum_l \frac{ c_{l}^{2} (\nu_{l}+\frac{2l+1}{2})}{1-c_{l}^{2} x_{i}}-{\sum_{j\neq i}{\frac{2}{x_{i}-x_{j}}}} \,.
\end{equation}

A similar structure to Eq.~\eqref{quasi} was first used by Gaudin~\cite{Gaudin76} as a guess in obtaining exact solutions of interaction systems, which is now confirmed to be a consistent operator form in composing the Bethe \emph{ansatz} wave function, Eq.~\eqref{wave} for the current tetraquark system.

Our formalism and methods for masses of heavy tetraquarks are the same as the procedure in Ref.~\cite{pan2006, aj181, aj182}. The representation~\eqref{a01} is totally symmetric, corresponding to the fact that the excitations (vibrations and rotations) are bosonic in nature. So, we have to define the number of bosons in our system. Here the boson number value depicts the total number of vibrational states in the representation $[N]$.

\begin{figure}[!t]
\includegraphics[width=0.475\textwidth]{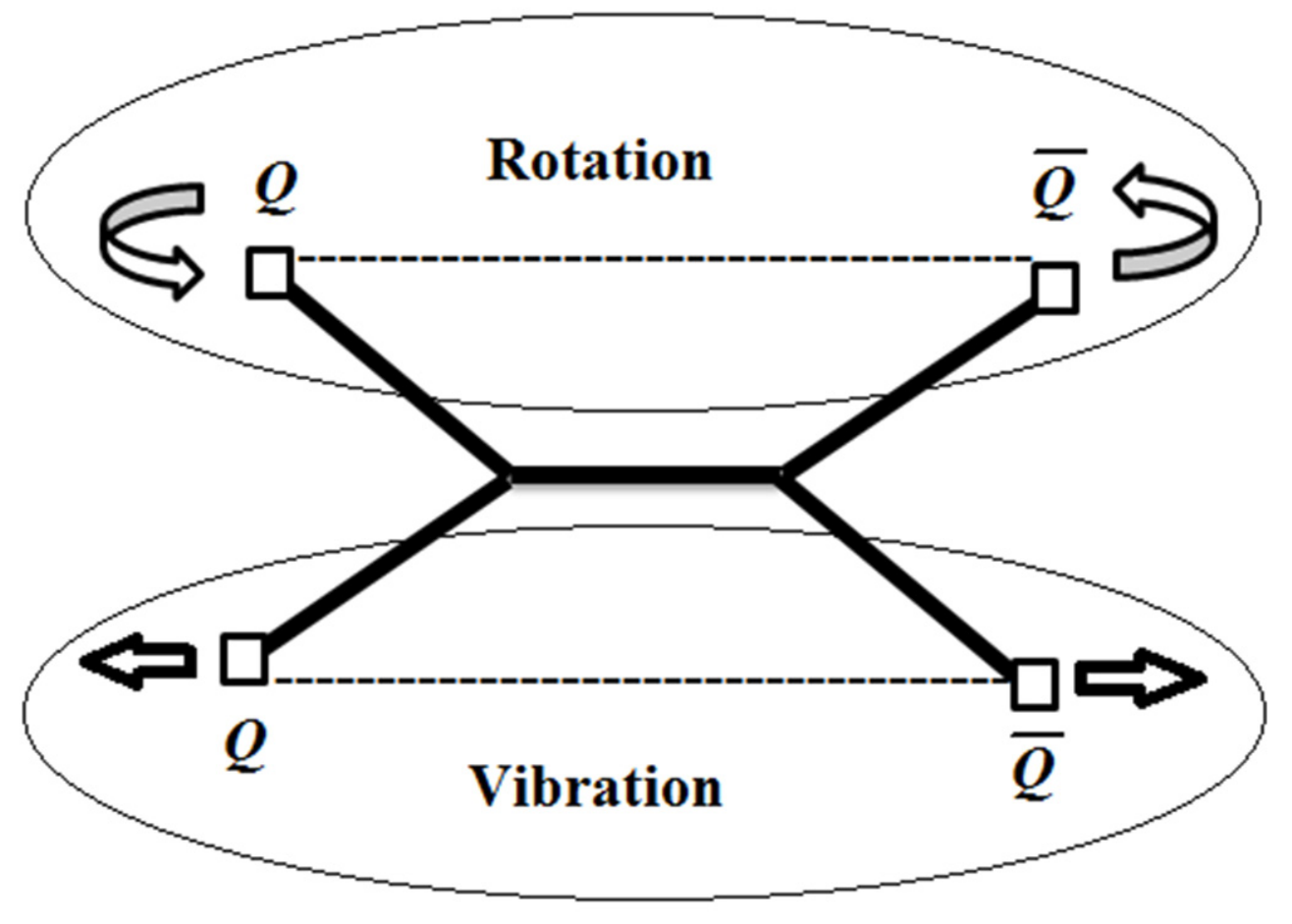}
\caption{\label{fig:rotvib} Schematic representation of the rotational and vibrational degrees of freedom in the studied tetraquark systems.}
\end{figure}

The quantum phase transition occurs between vibrational and rotational limits in the full heavy tetraquark pairing model. The quark (antiquark) configuration can perform vibrations and rotations (Fig.~\ref{fig:rotvib}) defined by the quantum numbers $\nu_{Q_i}$, $\nu_{\bar{Q}_i}$ and $J$. We do not study here bending and twisting of tetraquarks since these are required to lie at higher masses. The pure configuration problem of Fig.~\ref{fig:rotvib} is slightly complicated by the fact that quarks and antiquarks have \emph{internal} degrees of freedom. Here we apply the solvable model to consider both geometric and internal excitations of the tetraquark masses. Our formalism and methods for masses of heavy tetraquarks are the same as the procedure in Ref.~\cite{pan2006}.

\section{Results}
\label{sec:results}

In the diquark--anti-diquark pairing model, the tetraquark mass can be determined by solving the eigenvalue problem of Eq.~(\ref{a01}). Moreover, the quantum numbers that define a tetraquark state are the spins of diquark and antidiquark clusters, and the total spin, spatial inversion symmetry and charge conjugation of the system, \emph{i.e.} the $J^{PC}$ quantum numbers. Following Ref.~\cite{yang}, for a $Q_1Q_2\bar{Q}_3\bar{Q}_4$ system, the quantum labels are $J^{PC}$=$0^{++}$, $1^{+-}$ and $2^{++}$, and thus we have:
\begin{enumerate}
\item Two states for the scalar system:
\begin{subequations}
\begin{align}
|0^{++}\rangle &= |0_{Q_1Q_2}, 0_{\bar Q_3 \bar Q_4};J=0\rangle \,, \\
|0^{++\prime}\rangle &= |1_{Q_1Q_2}, 1_{\bar Q_3 \bar Q_4};J=0\rangle \,.
\label{eq:zeropp}
\end{align}
\end{subequations}
\item Three states for the vector system:
\begin{subequations}
\begin{align}
|A\rangle &= |0_{Q_1Q_2}, 1_{\bar Q_3 \bar Q_4};J=1\rangle \,, \\
|B\rangle &= |1_{Q_1Q_2}, 0_{\bar Q_3\bar Q_4};J=1\rangle \,, \\
|C\rangle &= |1_{Q_1Q_2}, 1_{\bar Q_3\bar Q_4};J=1\rangle \,.
\label{eq:onep}
\end{align}
\end{subequations}
Under charge conjugation, we have different configurations in which $|A\rangle$ and $|B\rangle$ interchange while $|C\rangle$ is odd. Thus, the $J^P=1^+$ involves one $C$-even and two $C$-odd states:
\begin{subequations}
\begin{align}
|1^{++}\rangle &= \frac{1}{\sqrt{2}}(|A\rangle+|B\rangle) \,, \\
|1^{+-}\rangle &= \frac{1}{\sqrt{2}}(|A\rangle-|B\rangle) \,, \\
|1^{+-\prime}\rangle &= |C\rangle \,.
\label{eq:onepp}
\end{align}
\end{subequations}
Note here that we must select the appropriate values for the spin of $Q_1{\bar Q_3}$ and $Q_2\bar Q_4$. This means that the only state with $C=+$ is that in which $Q_1\bar{Q}_3$ has spin $S_{Q_1\bar{Q}_3}=1$.
\item One state for the tensor system:
\begin{equation}
|2^{++}\rangle = |1_{Q_1Q_2}, 1_{\bar Q_3 \bar Q_4}; J=2\rangle \,,
\label{eq:twopp}
\end{equation}
where this state has also $S_{Q_1\bar Q_3}=1$.
\end{enumerate}


\subsection{The $[cc][\bar{c}\bar{c}]$ system}

In the pairing tetraquark model, the rigid and non-rigid phases correspond, respectively, to the $SO(10)$ and $U(9)$ symmetry cases. Both are idealized situations and must coexist in real world. Therefore, the $U(9) \leftrightarrow SO(10)$ transitional region is where the two phases coexist and vibrational-rotational modes appear. 

The parameters in the transitional region are called the phase parameters since the $c_{Q_i}=1$, with $i=1,\ldots,4$, case corresponds to the rotational mode, while the $c_{Q_i}=0$ case corresponds to the vibrational mode. We first can calculate the mass spectrum of the pairing tetraquark model with fixed phase parameters. Then, the transitional spectra from one phase to the other can be obtained modifying the phase parameters within the closed interval $[0,1]$.

From a transitional theory point of view, the ideal way of extracting the values of the phase coefficients is looking at the meson-meson thresholds $\eta_{c}(1S)\eta_{c}(1S)$ and $J/\psi(1S)J/\psi(1S)$ for $J^{PC}=0^{++}$,  $\eta_{c}(1S)J/\psi(1S)$ for $J^{PC}=1^{+-}$, and $J/\psi(1S)J/\psi(1S)$ for $J^{PC}=2^{++}$. Our values are $c_{Q_1}=0.92$, $c_{Q_2}=1$, and $c_{\bar{Q}_3}=c_{\bar{Q}_4}=0$, which results into the following masses
\begin{align}
|0^{++\prime}\rangle &= |1_{cc}, 1_{\bar c \bar c};J=0\rangle: M=5.978\,\text{GeV} \,, \\
|1^{+-\prime}\rangle &= |1_{cc}, 1_{\bar c\bar c};J=1\rangle: M=6.155\,\text{GeV} \,, \\
|2^{++}\rangle &= |1_{cc}, 1_{\bar c \bar c}; J=2\rangle: M=6.263\,\text{GeV} \,,
\end{align}
for the $T_{4c}$ tetraquark system.


\subsection{The $[bb][\bar{b}\bar{b}]$ system}

The situation here is very similar with respect the case above. This time, from a transitional theory point of view, the extraction phase coefficients must be performed attending to the meson-meson thresholds $\eta_{b}(1S)\eta_{b}(1S)$ and $\Upsilon(1S)\Upsilon(1S)$ for $J^{PC}=0^{++}$, $\eta_{b}(1S)\Upsilon(1S)$ for $J^{PC}=1^{+-}$, and $\Upsilon(1S)\Upsilon(1S)$ for $J^{PC}=2^{++}$. Our numerical values are $c_{Q_1}=0.97$, $c_{Q_2}=1$, $c_{\bar{Q}_3}=1$ and $c_{\bar{Q}_4}=0$, which provide the following masses:
\begin{align}
|0^{++\prime}\rangle &= |1_{bb}, 1_{\bar b \bar b};J=0\rangle: M=18.752\,\text{GeV} \,, \\
|1^{+-\prime}\rangle &= |1_{bb}, 1_{\bar b\bar b};J=1\rangle: M=18.805\,\text{GeV} \,, \\
|2^{++}\rangle &= |1_{bb}, 1_{\bar b \bar b}; J=2\rangle: M=18.920\,\text{GeV} \,,
\end{align}
for the $T_{4b}$ tetraquark system.


\subsection{The $[bc][\bar{b}\bar{c}]$ system}

The final structure analyzed in this work is the $T_{2bc}=[bc][\bar{b}\bar{c}]$ tetraquark system. In this case, the $[bc]$ diquark spin can be either $0$ or $1$ and thus all states analyzed at the beginning of this section are possible. Again, from a transitional theory point of view, the best extraction procedure of the control parameters in the $T_{2bc}$ tetraquarks are the corresponding meson-meson families which deliver the following values: $c_{Q_1}=c_{Q_2}=1$, and $c_{\bar{Q}_3}=c_{\bar{Q}_4}=0$. The masses computed in this case can be classified as follows:
\begin{itemize}
\item[(i)] The $J^{PC}=0^{++}$ contains two scalar states with masses
\begin{align}
|0^{++}\rangle &= |0_{bc}, 0_{\bar b \bar c};J=0\rangle: M=12.359\,\text{GeV} \,, \\
|0^{++\prime}\rangle &= |1_{bc}, 1_{\bar b \bar c};J=0\rangle: M=12.503\,\text{GeV} \,.
\end{align}
\item[(ii)] The $J^{PC}=1^{+-}$ contains two states with masses
\begin{align}
|1^{+-}\rangle &= \frac{1}{\sqrt{2}}(|0_{bc}, 1_{\bar b \bar c};J=1\rangle & \nonumber \\
&
-|1_{bc}, 0_{\bar b\bar c};J=1\rangle): M=12.896\,\text{GeV} \,, \\[1ex]
|1^{+-\prime}\rangle &= |1_{bc}, 1_{\bar b\bar c};J=1\rangle: M=12.016\,\text{GeV} \,.
\end{align}
\item[(iii)] The $J^{PC}=1^{++}$ contains one state with mass
\begin{align}
|1^{++}\rangle &= \frac{1}{\sqrt{2}}(|0_{bc}, 1_{\bar b \bar c};J=1\rangle \nonumber \\
&
+ |1_{bc}, 0_{\bar b\bar c};J=1\rangle): M=12.155\,\text{GeV} \,.
\end{align}
\item[(iv)] The $J^{PC}=2^{++}$ contains one state with mass
\begin{align}
|2^{++}\rangle = |1_{bc}, 1_{\bar b \bar c}; J=2\rangle: M=12.897\,\text{GeV} \,.
\end{align}
\end{itemize}


\section{Discussion}
\label{sec:discussion}

\begin{figure}[!t]
\includegraphics[width=0.475\textwidth]{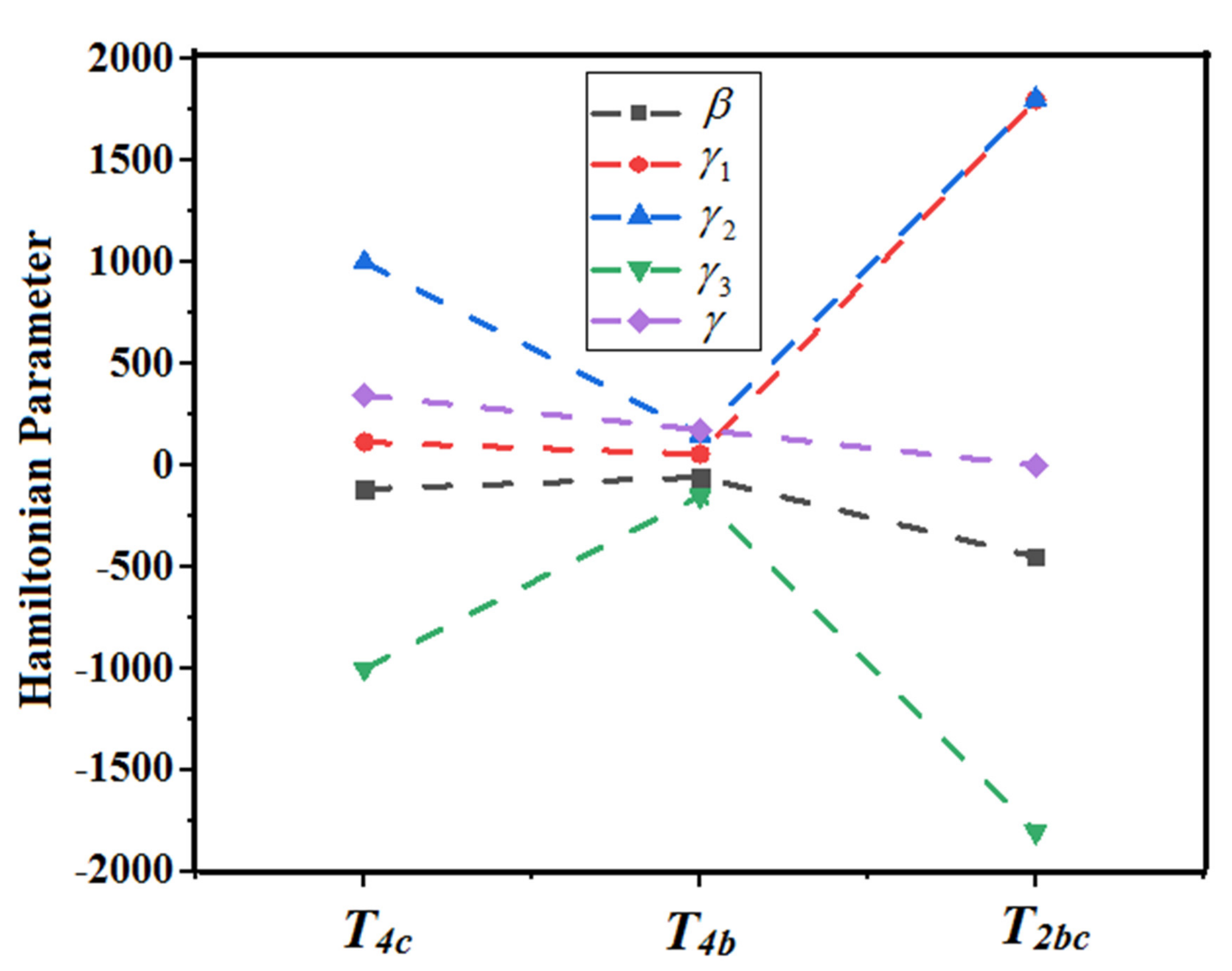}
\caption{\label{fig:parameters} The resulting parameters of the Hamiltonian when predicting the tetraquark masses based on the diquark-antidiquark pairing model. In the calculation, the effective $g$-factor is taken to be $1$.}
\end{figure}

\begin{table*}\label{t1}
\caption{\label{tab:QQmass} Masses of fully-heavy tetraquark systems as computed within the theoretical framework presented herein. The meson-meson threshold is $E_{th}$, the $\Delta=M-E_{th}$ represents the energy distance of the tetraquark with respected its lowest meson-pair threshold. Notation $s$ and $a$ indicates scalar and axial-vector diquarks.}
\begin{ruledtabular}
\begin{tabular}{ c c c c c c c }
Structure &Configuration   & $J^{PC}$ & $M_{tetra}$ in this work (GeV) & Threshold & $E_{th}$ (GeV)& $\Delta$ (GeV)
\\
\hline
\multirow{4}{*}{$T_{4c}$=$[cc] [\bar c \bar c]$} & \multirow{4}{*}{$a \bar a$} & \multirow{2}{*}{$0^{++}$} & \multirow{2}{*}{5.978} & $\eta_{c}(1S)\eta_{c}(1S)$ & 5.968 & 0.01
\\
\cline{5-7}
& & & &  $J/\psi(1S)J/\psi(1S)$ & 6.194 & -0.216
\\
\cline{3-7}
& & $1^{+-}$ & 6.155 & $\eta_{c}(1S)J/\psi(1S)$ & 6.081 & 0.074
\\
\cline{3-7}
& & $2^{++}$ & 6.263 & $J/\psi(1S)J/\psi(1S)$ & 6.194 & 0.069
\\
\cline{1-7}
\multirow{4}{*}{$T_{4b}$=$[bb][ \bar b \bar b]$} & \multirow{4}{*}{$a \bar a$} & \multirow{2}{*}{$0^{++}$} & \multirow{2}{*}{18.752} & $\eta_{b}(1S)\eta_{b}(1S)$ & 18.797 & -0.045
\\
\cline{5-7}
  & & & &  $\Upsilon(1S)\Upsilon(1S)$ & 18.920 & -0.168
\\
\cline{3-7}
  & & $1^{+-}$ & 18.808 & $\eta_{b}(1S)\Upsilon(1S)$ & 18.859 & -0.051
\\
\cline{3-7}
  & & $2^{++}$ & 18.920 & $\Upsilon(1S)\Upsilon(1S)$ & 18.920 & 0.0
  \\
  \cline{1-7}
  \multirow{21}{*}{$T_{2bc}$=$[bc][ \bar b \bar c]$} & \multirow{10}{*}{$a \bar a$} & \multirow{4}{*}{$0^{++}$} & \multirow{4}{*}{12.503} & $\eta_{b}(1S)\eta_{c}(1S)$ & 12.383 & 0.12
\\
\cline{5-7}
  & & & & $J/\psi(1S)\Upsilon(1S)$ & 12.557 & -0.054
\\
\cline{5-7}
  & & & & $B_{c}^{\pm}  B_{c}^{\mp}$ & 12.550 & -0.047
\\
\cline{5-7}
  & & & & $B_{c}^{*\pm} B_{c}^{*\mp}$ & 12.666 & -0.163
\\
\cline{3-7}
  & & \multirow{4}{*}{$1^{+-}$} & \multirow{4}{*}{12.016} & $\eta_{c}(1S)\Upsilon(1S)$ & 12.444 &-0.428
\\
\cline{5-7}
  & & & & $J/\psi(1S)\eta_{b}(1S)$ & 12.496 & -0.48
\\
\cline{5-7}
  & & & & $B_{c}^{\pm}  B_{c}^{*\mp}$ & 12.608 & -0.592
\\
\cline{5-7}
  & & & & $B_{c}^{* \pm}  B_{c}^{* \mp}$ & 12.666 & -0.65
\\
\cline{3-7}
  & & \multirow{2}{*}{$2^{++}$} & \multirow{2}{*}{12.897} & $J/\psi(1S)\Upsilon(1S)$ & 12.557 & 0.34
\\
\cline{5-7}
  & & & & $B_{c}^{* \pm}  B_{c}^{*\mp}$ & 12.666 & 0.231
\\
\cline{2-7}
  & \multirow{7}{*}{$\frac{1}{\sqrt{2}}(a \bar s \pm s \bar a)$} & \multirow{3}{*}{$1^{++}$} & \multirow{4}{*}{12.155} & $J/\psi(1S)\Upsilon(1S)$ & 12.557 & -0.402
\\
\cline{5-7}
  & & & & $B_{c}^{\pm} B_{c}^{*\mp}$ & 12.608 & -0.453
\\
\cline{5-7}
  & & & & $B_{c}^{*\pm}  B_{c}^{*\mp}$ & 12.666 & -0.511
\\
\cline{3-7}
  & & \multirow{4}{*}{$1^{+-}$} &  \multirow{4}{*}{12.896} & $\eta_{c}(1S)\Upsilon(1S)$ & 12.444 & 0.452
\\
\cline{5-7}
  & & & & $J/\psi(1S)\eta_{b}(1S)$ & 12.496 & 0.4
\\
\cline{5-7}
  & & & & $B_{c}^{\pm} B_{c}^{*\mp}$ & 12.608 & 0.288
\\
\cline{5-7}
  & & & & $B_{c}^{*\pm}  B_{c}^{*\mp}$ & 12.666 & 0.23
\\
\cline{2-7}
  & \multirow{4}{*}{$s \bar s$} & \multirow{4}{*}{$0^{++}$} & \multirow{4}{*}{12.359} & $\eta_{c}(1S)\eta_{b}(1S)$ & 12.383 & -0.024
\\
\cline{5-7}
  & & & & $J/\psi(1S)\Upsilon(1S)$ & 12.557 & -0.198
\\
\cline{5-7}
  & & & & $B_{c}^{\pm}  B_{c}^{\mp}$ & 12.550 & -0.191
\\
\cline{5-7}
  & & & & $B_{c}^{*\pm}  B_{c}^{*\mp}$ & 12.666 & -0.307
\\
\end{tabular}
\end{ruledtabular}
\end{table*}

\begin{table*}\label{t2}
\caption{\label{tab:cm1} Comparison of our results with theoretical predictions for the masses of $T_{4b}=[bb][\bar{b}\bar{b}]$, and $T_{4c}=[cc][\bar{c}\bar{c}]$ tetraquarks. All results are in GeV.}
\begin{ruledtabular}
\begin{tabular}{ccccccc}
Reference & \multicolumn{3}{c}{$bb \bar b \bar b$}& \multicolumn{3}{c}{$cc \bar c \bar c$}\\
\cline{2-4} \cline{5-7} & $0^{++}$ & $1^{+-}$ & $2^{++}$ & $0^{++}$ & $1^{+-}$ & $2^{++}$\\
\hline
 \centering{This paper} & 18.752 & 18.808 & 18.920& 5.978 & 6.155 & 6.263  \\
 \centering{\cite{p2}} & 18.460-18.490 & 18.320-18.540 & 18.320-18.530& 6.460-6.470  & 6.370-6.510 & 6.370-6.510  \\
  \centering{\cite{FullBeauty2019}}& 18.690  &- &-&-&-&- \\
  \centering{\cite{FullHeavy2019sec}} & 18.748 & 18.828 & 18.900& 5.883 & 6.120 & 6.246 \\
  \centering{\cite{FullHeavy2018}} & 18.750 &- &-&$ <6.140$ &- &-   \\
  \centering{\cite{D12},\cite{blln}}  & 18.754 & 18.808 & 18.916& 5.966 & 6.051 & 6.223  \\
 \centering{\cite{FullHeavy2017,Karliner:2020dta}} & $18.826 $ &- & $18.956$& $6.192$ & -&$6.429$ \\
 \centering{\cite{SumR2,E18}}& $18.840 $ & $18.840 $ & $18.850 $ & $5.990 $ & $6.050 $ & $6.090 $   \\
 \centering{\cite{Chen}}&19.178&19.226&19.236&-&-&-\\
 \centering{\cite{Jin:2020jfc}}&19.237&19.264&19.279&6.314&6.375&6.407\\
\centering{\cite{WLZ}}&19.247 &19.247  &19.249&6.425&6.425&6.432 \\
\centering{\cite{FullHeavy2019,liu:2020eha}} & 19.322 & 19.329 & 19.341& 6.487 & 6.500 & 6.524  \\
\centering{\cite{p5}}&19.329&19.373&19.387&6.407&6.463&6.486\\
 \centering{\cite{Lu:2020cns}}&19.255&19.251&19.262&6.542&6.515&6.543\\
  \centering{\cite{p1}} & 20.155 & 20.212 & 20.243& 6.797 & 6.899 & 6.956 \\
  \centering{\cite{FullCharm2017,E19}} &-&-&-& 5.969 & 6.021 & 6.115\\
  \centering{\cite{102}} &-&-&-& 6.695 & 6.528 & 6.573\\
  \centering{\cite{103}} &-&-&-& 6.480 & 6.508 & 6.565\\
  \centering{\cite{tetrababc}} & 19.666 & 19.673 & 19.680& 6.322 & 6.354 & 6.385 \\
   \centering{\cite{tetrac}} & - & - & -& 6.510 & 6.600 & 6.708 \\
   \centering{\cite{25}} & 18.981 & 18.969 &19.000& 6.271 & 6.231 & 6.287 \\
   \centering{\cite{26}} & 19.314 &19.320 &19.330& 6.190 & 6.271 &6.367 \\
   \centering{\cite{p4}}~set. I & 18.723 & 18.738 &20.243& 5.960 & 6.009 & 6.100 \\
   \centering{\cite{p4}}~set. II & 18.754 & 18.768 &18.797& 6.198 & 6.246 & 6.323 \\
   \centering{\cite{zha}} & 19.226 & 19.214 &19.232& 6.476 & 6.441 & 6.475 \\
  \end{tabular}
\end{ruledtabular}
\end{table*}

\begin{table*}\label{t3}
\caption{\label{tab:cm2} Comparison of our results with theoretical predictions for the masses of $T_{2bc}=[bc][\bar{b}\bar{c}]$  tetraquarks. All results are in GeV.}
\begin{ruledtabular}
\begin{tabular}{ccccccc}
\centering{Reference} & \multicolumn{3}{c}{{$a \bar a$}} & \multicolumn{2}{c}{{$\frac{1}{\sqrt{2}}(a \bar s \pm s \bar a)$}} & \multicolumn{1}{c}{{$s \bar s$}}
\\[1ex]
\cline{2-4} \cline{5-6} \cline{7-7}
 & \centering{$0^{++}$} & \centering{$1^{+-}$} & \centering{$2^{++}$} & \centering{$1^{++}$} & \centering{$1^{+-}$} & \multicolumn{1}{c}{$0^{++}$}
\\
\hline
\centering{This paper} & 12.503 & 12.016 & 12.897 & 12.155 & 12.896 & 12.359\\
  \centering{\cite{D12}} &12359 & 12424  & 12566 & 12485 &12488 & 12471\\
     \centering{\cite{FullHeavy2019sec}} & 12374 & 12491 & 12576 &12533 & 12533 & 12521\\
    \centering{\cite{FullHeavy2018}} & $<12620$ &- &- &- &- &-\\
   \centering{\cite{Chen2}}&12746&12804&12809&-&12776&-\\
   \centering{\cite{p5}}&12829&12881&12925&-&-&-\\
\centering{\cite{FullHeavy2019}} & 13035  & 13047 & 13070 & 13056 & 13052 & 13050\\
  \centering{\cite{p1}} & 13483 & 13520 & 13590 & 13510 & 13592 & 13553\\
\end{tabular}
\end{ruledtabular}
\end{table*}

In the calculation procedure, we fix the Hamiltonian parameters and allow the phase parameters to vary during the transition. In Ref.~\cite{f2} we showed that the quantum number of the amount of bosons can be taken in the $N\to \infty$ limit. Moreover, it was adequate to take $N$ large enough to cover all known and unknown states up to a maximum value of the quantum number of the angular momentum, and other quantum numbers connected with applications. In the present investigation, we take this to be the same as that used in Ref.~\cite{f2} with $N=100$. 

The trend of Hamiltonian is similar to that of the $O(4)$ limit condition proposed in mesons when the control parameter is taken to be $1$. Most importantly, our investigation shows that the control parameters $c_{\bar{Q}_3}$ and $c_{\bar{Q}_4}$ cannot be taken to be $1$ when heavy antiquarks are involved except for $T_{4b}$ tetraquarks. This is because the masses of $T_{4b}$ tetraquarks are $2-3$ times heavier than $T_{2bc}$ and $T_{4c}$ ones. In this condition for heavy mass tetraquarks, the effect of pairing strength is strong, which can be seen in the fact that $c_{Q_1}$ for $T_{2bc}$ is larger than in the $T_{4c}$ case.

The values of the parameters in Hamiltonian for the mentioned structures are given in Figure~\ref{fig:parameters}. In the transition region, $\alpha$ is taken to be $1.5$. Since the vibrational-rotational transition within the pairing model is a second-order quantum phase transition, the masses wave functions in the $U(9)$ model of studied tetraquarks behave smoothly with respect changes in the parameters, which allows us to fix them in the transition region.

Table~\ref{tab:QQmass} shows the difference between the calculated tetraquark masses and meson-pair threshold. We present the values of $\Delta=M_{tetra}-E_{th}$, where $M_{tetra}$ and $E_{th}$ are the tetraquark mass and its lowest meson-meson threshold, respectively. A negative $\Delta$ indicates that the tetraquark state lies below the threshold of the fall-apart decay into two mesons and consequently should be stable. Besides, a state with a small positive value for the $\Delta$ could also be observed as a resonance since the phase space would suppress its partial decay width. The remaining states, with large positive $\Delta$ values, are supposed to be broad and challenging to recognize in experimental analyses.

Our analysis confirms that the control parameter $c_{Q_1}$ deviating a little from $1$ appears better in the extraction of the tetraquark masses, specially for comprehensive $T_{2bc}$ families. One can also see that, in the $T_{4b}=[bb][\bar{b}\bar{b}]$ states, the higher contribution comes from the pairing of $c_{\bar{Q}_3}$ and $c_{\bar{Q}_4}$ quarks. This means that at high energy, around $18-19\,\text{GeV}$, phase parameters for $\bar{Q}_3$ and $\bar{Q}_4$ quarks begin to play an essential role in computing tetraquark masses; while in the low energy regime, there is a competition between the ${Q}_1$ and ${Q}_2$. 

According to the above definition, it can be claimed that the energy spectra of the studied fully-heavy tetraquarks in which $c_{Q_i} \sim 0.9-1.0$ corresponds to a rotational phase. Note also that a change of $\pm15\%$ in all coefficients produce a maximum variation of $30\%$, $23\%$, $17\%$ in a particular channel's mass of $T_{4c}$, $T_{4b}$ and $T_{2bc}$ tetraquark systems, respectively; having that all remaining masses experience lesser modifications.

Finally, the results obtain herein with the pairing model are compared with the prediction of previous theoretical calculations in Tables~\ref{tab:cm1} and~\ref{tab:cm2}. One can deduce that the theory fairly reproduces the other works, indicating that our solvable model could still play an essential role in the prediction of fully-heavy tetraquark mesons. In order to do so, a possible improvement is to include the large-$N$ limit of the pure pairing Hamiltonian to gain a better understanding of the multiquark dynamics.


\section{Summary}
\label{sec:summary}

Inspired by the problem of solving the interacting $sl$-boson system in the transitional region, the solvable extended Hamiltonian that includes multi-pair interactions has been considered to provide the mass spectra of fully-heavy tetraquarks. Numerical extractions of $T_{4c}$, $T_{4b}$, and $T_{2bc}$ ground state masses, within the algebraic model in which the Bethe \emph{ansatz} is adopted, were carried out to test the theory. The results reveal that the $U(9) \to SO(10)$ Hamiltonian could predict spectra in fair agreement with other theoretical approaches.

Finally, the solvable technique introduced in this manuscript may also be helpful in diagonalizing more general multiquark systems, which will be considered in future work.


\begin{acknowledgments}
We thank Prof. Xue-Qian Li for stimulating discussions on the tetraquarks and his suggestion on doing this work.
This work has been partially funded by the National Natural Science Foundation of China (11875171, 11675071, 11747318), the  U.S. National Science Foundation (OIA-1738287 and ACI -1713690), {U. S. Department of Energy (DE-SC0005248)}, the Southeastern Universities Research Association, the China--U.S. Theory Institute for Physics with Exotic Nuclei (CUSTIPEN) (DE-SC0009971), and the LSU--LNNU joint research program (9961) is acknowledged). The Ministerio Espa\~nol de Ciencia e Innovaci\'on under grant no. PID2019-107844GB-C22; and Junta de Andaluc\'ia, contract nos. P18-FRJ-1132, Operativo FEDER Andaluc\'ia 2014-2020 UHU-1264517, and PAIDI FQM-370.
\end{acknowledgments}


\end{document}